# Not a Silver Bullet for Loneliness: How Attachment and Age Shape Intimacy with AI Companions

Raffaele Ciriello, University of Sydney, raffaele.ciriello@sydney.edu.au

Uri Gal, University of Sydney, uri.gal@sydney.edu.au

Ofir Turel, University of Melbourne, oturel@unimelb.edu.au

*Abstract*

Artificial intelligence (AI) companions are increasingly promoted as solutions for loneliness, often overlooking how personal dispositions and life-stage conditions shape artificial intimacy. Because intimacy is a primary coping mechanism for loneliness that varies by attachment style and age, we examine how different types of users form intimate relationships with AI companions in response to loneliness. Drawing on a hermeneutic literature review and a survey of 277 active AI companion users, we develop and test a model in which loneliness predicts intimacy, moderated by attachment insecurity and conditioned by age. Although the cross-sectional data limits causal inference, the results reveal a differentiated pattern. Loneliness is paradoxically associated with *reduced* intimacy for securely attached users but with increased intimacy for avoidant and ambivalent users, while anxious users show mixed effects. Older adults report higher intimacy even at lower loneliness levels. These findings challenge portrayals of AI companions as universal remedies for loneliness. Instead, artificial intimacy emerges as a sociotechnical process shaped by psychological dispositions and demographic conditions. The study clarifies who is most likely to form intimate relationships with AI companions and highlights ethical risks in commercial models that may capitalise on user vulnerability.

# 1. INTRODUCTION

Loneliness has reached epidemic levels, with governments and the World Health Organization warning of its elevated risks of anxiety, depression, dementia, suicide, and cardiovascular disease (Ernst et al. 2022; Statista 2025; WHO 2023). Although loneliness is a relational and systemic phenomenon (Hughes et al. 2024), public and commercial discourses often present it as an individual affliction amenable to technocratic remedies (De Freitas 2024; Hughes 2024). Yet, unlike chosen solitude, loneliness reflects unmet needs for meaningful connection shaped by both personal dispositions and social conditions (Cacioppo et al. 2015; Mund et al. 2020; Vanhalst et al. 2015). Digital technologies are thus double-edged: they can facilitate perceived responsiveness (De Freitas et al. 2025; Park & Kim 2022; Sullivan et al. 2023), yet also fragment sociality and foster maladaptive dependence (Al-Obaydi & Pikhart 2025; Jacobs 2024; Turel & Serenko 2012).

This ambivalence is evident in the rapid rise of *AI companions*, here defined as generative conversational agents designed and used to simulate human-like intimate relationships. Often marketed as remedies for loneliness (Chen et al. 2024), and increasingly studied as such (De Freitas et al. 2025; Sullivan et al. 2023), AI companions have also been linked to intensified loneliness and problematic use among vulnerable individuals (Alabed et al. 2024; Herbener & Damholdt 2025; Hu & Thomas 2025; Jacobs 2024). The relational dynamics that underpin these divergent outcomes remain poorly understood.

Recent Information Systems (IS) research has examined affective aspects of human–AI interaction, focusing on how emotional cues in chatbots enhance service quality, consumer trust, and communication (Chandra et al. 2022; Han et al. 2023; Meier et al. 2024; Saffarizadeh et al. 2024). While valuable for understanding business outcomes, this literature offers limited insight into how AI companions may alleviate loneliness or create new harms across different user groups. As the societal impacts of AI become increasingly contested (Sabherwal & Grover 2024), artificial intimacy emerges as a critical yet understudied construct (Brooks 2024).

*Intimacy* is an interpersonal process of affective disclosure and perceived responsiveness (Laurenceau et al. 1998; Reis & Shaver 1988). It is central to meaningful relationships and a primary coping mechanism for loneliness (Cacioppo et al. 2015; Mund et al. 2020). Many AI companions are explicitly designed to simulate intimacy without genuine social presence (Janssen et al. 2014; Xie et al. 2023; Potdevin et al. 2021). Even general-purpose systems such as ChatGPT, not originally designed for intimacy, are now primarily used for therapy and companionship (Zao-Sanders 2025). Yet simulated intimacy carries risks of distorted judgement, emotional manipulation, and dependency (Chen et al. 2025; Gersen 2019; Schroeder et al. 2017; Su et al. 2019). Understanding when and for whom intimacy emerges is thus essential for evaluating benefits and harms.

Because both loneliness and intimacy are intertwined with psychological dispositions and social conditions (Hughes et al. 2024; Mund et al. 2020; Reis & Shaver 1988), it is implausible that intimacy with AI companions develops uniformly. Loneliness varies across the life span and personality traits (Mund et al. 2020; von Soest et al. 2020), and technology's effects are conditioned by psychological dispositions (Jin 2013; Phu & Gow 2019). Although prior work shows that individual differences shape loneliness in digitally mediated contexts (Balki et al. 2022; Mund et al. 2020; Pittman & Reich 2016), the mechanisms underlying how varying levels of loneliness may affect the formation of intimacy with AI companions remain understudied.

Attachment theory provides a pertinent framework for examining such variability because it predicts systematic differences in how individuals develop and experience intimacy (Ainsworth & Bowlby 1991; Hazan & Shaver 1987; Mikulincer 1998). According to the theory, internal models of the self and others shape orientations toward intimacy in patterned ways: secure individuals are comfortable with intimacy, avoidant individuals strive to minimise it and maintain distance, anxious individuals seek it excessively, and ambivalent individuals oscillate between desire and withdrawal (Bartholomew & Horowitz 1991; Feeney & Noller 1990; Spielmann et al. 2013). These dispositions shape how people cope with unmet relational needs (Mikulincer 1998; Shaver et al. 2005) and may condition their propensity to form intimate bonds with AI companions. Because AI companions offer controllable intimacy without the reciprocity, friction, or risks of human relationships, they may activate attachment strategies differently. Attachment styles may thus amplify, attenuate, or redirect the link between loneliness and artificial intimacy. Age further conditions this relationship. Older adults face heightened risks of chronic isolation (Carstensen 2021; Mund et al. 2020; von Soest et al. 2020) and may derive particular value from assistive technologies in the context of global demographic ageing (Latikka et al. 2022; Park & Kim 2022; von Soest et al. 2020). We therefore ask: *How do attachment style and age shape users' propensity to form intimate relationships with AI companions to cope with loneliness?*

We address this question through a cross-sectional survey of 277 active AI companion users. Using validated measures, we examine how loneliness relates to intimacy and how attachment style and age moderate this association. Although cross-sectional data limit causal inference, the correlational patterns reveal meaningful differentiation. Loneliness is associated with *lower* intimacy among securely attached users, *higher* intimacy among avoidant and ambivalent users, and mixed effects for anxious users. Older adults report higher intimacy even at lower levels of loneliness. These results challenge the assumption that loneliness uniformly drives artificial intimacy and instead suggest that AI companions may resonate most with individuals whose attachment style complicates human relationships.

Our contribution to IS research on AI companions is twofold. First, we identify and explain an apparent paradox: loneliness does not consistently increase intimacy with AI companions, but does so differentially across attachment styles, diverging from typical human relational patterns. Avoidant users, who typically down-regulate intimacy (Spielmann et al. 2013), seem to deepen artificial intimacy when relationships appear controllable and safe. Ambivalent users appear to do so when AI offers adjustable closeness without interpersonal cost (Bartholomew & Horowitz 1991). Secure users appear to treat AI companions as supplemental tools rather than substitutive partners. These findings position AI companions as relational technologies whose effects depend on psychological dispositions, challenging deterministic narratives that portray AI companionship as uniformly beneficial (De Freitas et al. 2025; Park & Kim 2022; Sullivan et al. 2023) or uniformly harmful (Alabed et al. 2024; Herbener & Damholdt 2025; Hu & Thomas 2025; Jacobs 2024). They also call for research on privacy, trust, and design that accounts for dispositional heterogeneity in emotional disclosure (Cacioppo et al. 2015; Jin 2013; Mund et al. 2020; Pittman & Reich 2016). For practitioners, the results imply that onboarding, customisation, and safeguards should reflect diverse attachment styles, rather than one-size-fits-all models.

Second, the concentration of intimacy among insecurely attached and lonely users raises ethical concerns about vulnerability in an emerging "intimacy economy" (Bozdag 2025). Insecure attachment is associated with distress, adverse health outcomes, and poorer relational functioning (Berant et al. 2001; Hardy & Barkham 1994; Maunder & Hunter 2001; Pietromonaco & Beck 2019). AI companions may thus disproportionately expose vulnerable users to manipulation, commercial exploitation, and privacy risks (Belk 2022; Gersen 2019; Su et al. 2019). As general-purpose chatbots are increasingly used for companionship (Zao-Sanders 2025), these findings highlight the need for duty-of-care obligations and safeguards against manipulative design.

The remainder of the paper reviews relevant literature, outlines the methodology, presents results, and discusses implications for IS scholarship and practice.

## 2. LITERATURE REVIEW

This section theorises how loneliness and attachment shape intimacy with AI companions. Loneliness is an aversive relational state that arises from unmet needs for meaningful connection. Intimacy is a core coping mechanism through which such needs are typically addressed, characterised by affective disclosure and perceived responsiveness, which AI companions are designed to simulate. Attachment theory explains systematic differences in individuals' propensity to seek intimacy, regulate closeness, and accept non-human relational substitutes. We conceptualise loneliness as antecedent, reflecting its role as a motivational state that may orient users toward AI companions as sources of connection. Intimacy constitutes the outcome, capturing the extent to which users experience the meaningful relational engagement these technologies promise. Attachment operates as the moderator, shaping how individuals interpret and respond to relational opportunities. The same level of loneliness may therefore prompt differently attached individuals to pursue distinct degrees of intimacy with AI companions. This positioning moves beyond assumptions of uniform benefit and enables the derivation of differentiated hypotheses about how artificial intimacy develops.

The following draws on a hermeneutic literature review conducted to interrogate assumptions shaping academic and commercial discourse on AI companions. Unlike systematic reviews that prioritise exhaustive coverage, hermeneutic reviews emphasise iterative interpretation and conceptual refinement, making them well suited to emergent, cross-disciplinary domains characterised by theoretical fragmentation (Boell & Cecez-Kecmanovic 2014). Research on AI companions spans psychology, IS, HCI, media studies, and communication, yet lacks an integrated framework linking loneliness, intimacy, and attachment. A hermeneutic approach enables abductive movement between texts, emerging themes, and theoretical development (Sarker et al. 2018). We engaged in iterative search–reading–interpretation cycles across Scopus, Google Scholar, and key disciplinary outlets (Boell & Cecez-Kecmanovic 2014). Search terms included *loneliness*, *intimacy*, *attachment*, combined with *artificial intelligence*, *chatbot*, *AI companion*, *virtual agent*, *relationship*, *companionship*, and related variants. Given the rapid evolution of the field, we included peer-reviewed conference papers and high-quality mainstream media analyses.

Three recurring assumptions emerged: 1) loneliness is often portrayed as if it affected everyone equally and thus treated as the primary driver of AI companion adoption and use; 2) intimacy (if considered at all) is often conceptualised as a static outcome rather than a dynamic interpersonal process shaped by sociotechnical mediation; and 3) attachment theory, although central to human relational variability, remains scarcely applied to human-AI companionship. These insights informed our conceptual model and guided the hypotheses linking loneliness, intimacy, and attachment.

### 2.1. Loneliness

Loneliness has intensified into a global public health concern. National surveys indicate that between one-third and, in some contexts, nearly half of adults report insufficient meaningful connection (Statista 2024a, 2024b, 2025), with associated health risks comparable to smoking and obesity (Ernst et al. 2022; WHO 2023). Yet loneliness is not uniform. Unlike chosen solitude, which can be restorative, loneliness reflects an unwanted discrepancy between desired and actual connection (Cacioppo et al. 2015; Vanhalst et al. 2015). It encompasses emotional loneliness (absence of an attachment figure), social loneliness (deficits in broader networks), and collective loneliness (disconnection from shared identities) (Cacioppo et al. 2015; Mund et al. 2020).

Loneliness also varies across the life course. It fluctuates more strongly in young adulthood as identity formation and shifting peer networks unsettle relational ties. In later life, shrinking time horizons lead individuals to prioritise emotionally meaningful, low-effort connections, even as reduced networks and community participation elevate the risk of chronic isolation (Carstensen 2021; Mund et al. 2020; von Soest et al. 2020). These age-related differences correspond to shifting motivations and capabilities for technology use (Tams 2022). Personality further shapes susceptibility, with extraversion buffering and neuroticism magnifying loneliness (von Soest et al. 2020). Such heterogeneity complicates portrayals of loneliness as a simple individual deficit amenable to technological fixes.

Digital technologies increasingly shape how loneliness is expressed and managed, yet their effects remain ambivalent. Social media affords low-cost contact (Phu & Gow 2019) and can support adolescents during crises (Cauberghe et al. 2020), while self-help communities may alleviate distress (Latikka et al. 2022). However, benefits are unevenly distributed. Individuals high in loneliness often gain little from online engagement (Jin 2013), and digital interactions often amplify underlying relational inequities (Phu & Gow 2019). Using social media to cope with stress can exacerbate it and prolong loneliness (Ghanayem et al. 2024). Persistent checking behaviours among lonely smartphone users further entrench dependency without delivering meaningful connection (Kim 2018). Reinforcement through social feedback, such as 'likes', may heighten addictive tendencies (Turel & Serenko 2012). Digital technologies therefore do not uniformly reduce loneliness but may deepen disparities in relational capacity.

These disparities were further intensified during the COVID-19 pandemic, which accelerated reliance on digitally mediated interaction while simultaneously disrupting in-person social infrastructures. Remote work, reduced communal activity, and financial strain compounded pre-existing inequities (Bruce et al. 2022; Camara et al. 2023; Ernst et al. 2022; Hajek & König 2020; Haucke et al. 2022). Although remote work enhances autonomy, it is associated with elevated loneliness (Cheng et al. 2023). More broadly, as digital services expand across healthcare, education, commerce, and entertainment, structural conditions conducive to chronic loneliness may deepen (Balki et al. 2022; Chen et al. 2024).

Against this backdrop, AI companions have emerged not merely as temporary substitutes, but as new intermediaries for managing relational needs (Maples et al. 2024). Commercial providers often market them as technological fixes for loneliness (Chen et al. 2023), with substantial uptake: companionship and therapy are now the most common uses of generative AI, including general-purpose systems not explicitly marketed as companions, such as ChatGPT (Zao-Sanders 2025). Early academic studies echo this narrative, often positioning AI companions as a remedy for loneliness (De Freitas et al. 2025; Jacobs 2024; Sullivan et al. 2023). Short-term chatbot interactions can indeed reduce loneliness and anxiety among university students (Kim et al. 2025), one-week engagement with AI companions can alleviate loneliness at levels comparable to interacting with another person (De Freitas et al. 2025), and frequent use of AI-enabled smart speakers reduces loneliness among older adults living alone (Park & Kim 2022). These findings support cautious optimism that AI companions may offer emotional support for some groups in some situations.

However, a parallel literature tempers this optimism. For chronically lonely users, AI companionship may foster maladaptive reliance, particularly when social anxiety, rumination, or anthropomorphism shape engagement (Hu et al. 2023). Adolescents who turn to chatbots report higher loneliness and lower perceived support than non-users (Herbener & Damholdt 2025). Studies describe relational trajectories in which AI companions become central to emotion regulation and identity work, sometimes displacing human relationships (Alabed et al. 2024). Critical accounts argue that AI companions may merely 'digitise' rather than resolve loneliness by offering simulated recognition without restoring social integration (Jacobs 2024). Rather than eliminating loneliness, AI companionship may precipitate a relational configuration that provides short-term emotional relief while reinforcing isolation in the long term.

In light of this mixed evidence, we reframe loneliness not as an outcome of AI companionship but as its antecedent. Loneliness may motivate intimacy-seeking with digital technology, but this drive is shaped by users' personal dispositions. Whether AI companions mitigate or magnify loneliness thus depends less on the technology itself than on how it is used and by whom. Despite growing research on human-AI relationships, no studies have examined how loneliness shapes the development of intimacy with AI companions or how individual differences condition this process. Because intimacy is a core coping mechanism for alleviating loneliness, it is important to understand how intimacy forms when the partner is artificial.

## 2.2. Intimacy

Rising loneliness reflects not only reduced social contact but a deficit in meaningful relationships. *Intimacy* is the relational process through which such meaning is constructed through cycles of affective self-disclosure and perceived responsiveness that create a sense of being known and cared for (Laurenceau et al. 1998; Reis & Shaver 1988). Sternberg (1986) identifies intimacy as a core component of meaningful relationships, and it is widely regarded as a primary coping mechanism through which loneliness is alleviated (Cacioppo et al. 2015; Mund et al. 2020). Because loneliness signals unmet needs for meaningful connection, it is important to understand how digital technologies shape how intimacy develops.

Digital technologies have not merely mediated intimacy but reconfigured it. Social media enables dispersed disclosure, fostering networked intimacy in which tagging, audience awareness, and indirect references blur public-private boundaries (Farci et al. 2017). Ephemeral, image-based platforms such as Snapchat or Instagram can heighten perceived intimacy while also introducing risks of overexposure (Kofoed & Larsen 2016; Pittman & Reich 2016). Offline relationships are similarly affected: phubbing – checking one's phone during in-person conversation – reduces partner intimacy even when noticed only peripherally (Vanden Abeele et al. 2019).

Commercial logics increasingly incentivise designing for intimacy. The quantity and valence of Facebook self-disclosure predict perceived intimacy (Park et al. 2011), and emotional gratifications foster intimacy, reinforcing platform loyalty (Lin & Chu 2021), whereas low perceived intimacy predicts lurking (Rau et al. 2008). In digital services, intimacy predicts continued use more strongly than perceived usefulness (Lee & Kwon 2011), and intimacy with frontline employees enhances customer loyalty (Yim et al. 2008). The exposure of conversational bots on dating platform Ashley Madison revealed how technological mediation unsettles assumptions that intimacy presupposes humanness (Harrison 2019). Digitally mediated intimacy thus becomes a sociotechnical accomplishment rather than a purely interpersonal one.

These developments reflect a broader transition from an attention economy toward an *intimacy economy*, where emotional and personal data are exchanged for personalised, AI-mediated experiences (Bozdag 2025). Emotional cues and self-disclosure of artificial agents can enhance trust and service quality (Han et al. 2023; Saffarizadeh et al. 2024). Anthropomorphic cues in smart assistants increase perceived intimacy and interaction satisfaction (Xie et al. 2023), and virtual agents displaying intimacy-related behaviours elicit emotional reactions independent of social presence (Potdevin et al. 2021). Adjacent ecosystems widen the intimate possibility space. Online cosplaying can foster intimacy (Chen et al. 2025), and interactive sex dolls may sustain romantic relationships beyond erotic gratification (Belk 2022; Su et al. 2019), challenging the assumption that intimacy requires reciprocity, consent, or personhood (Gersen 2019). Even before AI companions, intimacy had become commodified, commercialised, and increasingly decoupled from shared vulnerability.

Companionship provides the broader relational category within which intimacy may develop. Defined as sustained shared presence and mutual support (Rook 2014), *companionship* extends beyond humans to animals, media figures, robots, and dolls (Belk 2022; Giles 2002; Hughes et al. 2020). AI companions differ from sentient beings not only in their constant availability, but by simulating intimacy *as if* they were emotionally attuned despite lacking subjective experience (Chen et al. 2023). Because intimacy can foster sensitive disclosure (Cacioppo et al. 2015; Mund et al. 2020), it is important to understand how individuals experience intimacy with systems that mimic responsiveness while potentially aggregating private information at scale.

Individual differences shape digitally mediated intimacy. Among social media users, avoidant attachment predicts reduced intimacy, whereas attachment anxiety predicts heightened engagement (Wang 2025). Yet avoidant attachment can *increase* perceived intimacy when interactions carry no interpersonal risk (Spielmann et al. 2013), suggesting that some users may feel *more* intimate with artificial partners than with human ones.

These insights foreshadow the paradox demonstrated in our study: intimacy with AI companions may develop in the opposite direction to human relationships precisely because artificial partners alter the perceived risk structure of intimacy. Despite the rapid adoption of AI companions, no studies have examined whether attachment styles moderate intimacy with AI companions. This gap is consequential. If intimacy is the coping mechanism through which loneliness is alleviated, and if AI companions attract users whose relational needs stem from dispositional rather than situational factors, understanding how attachment shapes artificial intimacy becomes essential. Our study addresses this gap by theorising and empirically testing how attachment moderates the relationship between loneliness and intimacy with an AI companion.

### 2.3. Attachment

Attachment theory, originating in developmental psychology (Ainsworth & Bowlby 1991) and later extended to adult relationships (Hazan & Shaver 1987), explains enduring differences in caregiving, intimacy-seeking, and emotion regulation through internal models of the self and others (Mikulincer 1998). The literature commonly identifies four adult attachment styles: *secure* (low anxiety and avoidance), *avoidant* (emotional distancing and discomfort with intimacy), *anxious* (reassurance-seeking and fear of rejection), and *ambivalent* (a volatile combination of anxiety and avoidance, producing desire–withdrawal spirals) (Feeney & Noller 1990; Hazan & Shaver 1987). Population studies consistently find secure attachment as the most prevalent style (50–60 %), followed by avoidant (23–25 %), anxious (11–19 %), and a smaller ambivalent minority (Bakermans-Kranenburg & van Ijzendoorn 2009; Mickelson et al. 1997). The three non-secure styles are typically grouped as insecure attachment (Davies et al. 2009; De Carli et al. 2016).

Although attachment styles are not deterministic, they reliably predict differences in intimacy-seeking, perceived support, receptivity to relational substitutes (such as those offered by AI companions), as well as distinct cognitive-affective patterns. Secure attachment reflects confidence in others' availability, enabling constructive coping and reduced anger proneness (Mikulincer 1998; Millings et al. 2013). Avoidant attachment involves deactivation strategies that down-regulate intimacy, suppress affect, and minimise dependency (Spielmann et al. 2013), leading to reduced disclosure, lower expectations of responsiveness (Geller &

Bamberger 2009), and greater propensity in men to endorse hostile sexism (Hart et al. 2012). Anxious attachment involves hyperactivation characterised by rumination, heightened vigilance, and excessive reassurance-seeking (Shaver et al. 2005). For instance, anxiously attached individuals display elevated sexual motivation and dependency-driven motives for intimacy, whereas avoidance is associated with reduced emotional closeness and instrumental use of sex (Davis et al. 2004). Ambivalent attachment combines these tendencies in spirals where intimacy is alternately desired and feared (Hart et al. 2012). These mechanisms help explain why attachment style moderates intimacy in both offline and digitally mediated contexts, where relational cues may appear partial, ambiguous, or controllable.

Insecure attachment constitutes a major psychological vulnerability, associated with maladaptive emotion regulation, reduced well-being, and difficulty coping with distress (Maunder & Hunter 2001; Mikulincer 1998; Shaver et al. 2005). Longitudinal studies link attachment insecurity to poorer mental health, greater lifetime distress, and cognitive decline in old age (Berant et al. 2001; Hardy & Barkham 1994; Maunder & Hunter 2001). Meta-analytic evidence associates insecure attachment with heightened risk of depression and anxiety, interpersonal sensitivity, and chronic hypervigilance (Pietromonaco & Beck 2019). Depression pathways include dysfunctional attitudes, depleted self-esteem, and excessive reassurance-seeking (Roberts et al. 1996; Shaver et al. 2005). Compared with securely attached individuals, who are generally comfortable with intimacy (Feeney & Noller 1990), anxious individuals seek constant validation and show heightened dependency, whereas avoidant individuals suppress emotional needs and minimise support (Bartholomew & Horowitz 1991; Stanton et al. 2017). These vulnerabilities may also transmit intergenerationally, shaping susceptibility across family lines (Besser & Priel 2005). Insecurity further predicts hostility, anger dysregulation, and escapist coping (Mikulincer 1998), reduced responsiveness in parental caregiving (Millings et al. 2013), and poorer helping and feedback behaviour at work (Geller & Bamberger 2009; Hardy & Barkham 1994; Wu et al. 2014). These patterns shape how individuals cope with unmet relational needs and help explain why some users are more inclined to seek controllable relational substitutes, such as AI companions.

Emerging research shows that attachment styles also influence technology use. In romantic contexts, anxious attachment predicts more frequent social media use for reassurance and monitoring (Oldmeadow et al. 2013), including intrusive behaviours such as accessing a partner's private data without consent (Reed et al.

2015). Avoidant attachment is associated with reduced text-messaging but greater reliance on less immediate channels such as email, reflecting discomfort with emotional exposure (Morey et al. 2013). Sexting similarly varies by style: anxious individuals are more likely to initiate sexual messaging, whereas avoidant individuals sext more selectively and instrumentally (Drouin & Landgraff 2012; Weisskirch & Delevi 2011). Technology can also serve compensatory functions. People may form attachment-like bonds with mobile phones, particularly when anxiety is high (Konok et al. 2016), and romantic media consumption can act as a surrogate for intimacy (Mende et al. 2019). Meta-analytic evidence links insecure attachment to problematic internet use, suggesting greater reliance on digital technologies when interpersonal coping is constrained (Niu et al. 2023; Odaci & Çikrikçi 2014). Attachment also shapes trust in AI agents: experimentally increasing security cues elevates trust in AI, whereas activating anxiety cues reduces it (Gillath et al. 2021). In online dating, anxious attachment predicts lower perceived success and poorer affective outcomes (Hu & Thomas 2025).

These findings suggest that artificial intimacy may develop differently from human intimacy. Because AI companions offer consistency, controllability, and low interpersonal risk, insecurely attached individuals may find them easier to engage with than human partners. Securely attached individuals, by contrast, may not translate loneliness into artificial intimacy, treating AI companionship as supplemental rather than substitutive. Despite extensive evidence linking attachment to relationship functioning and technology use, no studies have examined how attachment moderates the association between loneliness and intimacy with AI companions. This gap is significant given the rapid expansion of companion systems and concerns that such technologies may attract, and potentially exploit, psychologically vulnerable users. Our study addresses this gap by theorising and empirically testing how attachment styles shape the link between loneliness and intimacy with an AI companion, including the surprising inversion we document relative to human–human relationships.

### 2.4. Literature Synthesis and Hypotheses Development

Prior research establishes that loneliness reflects unmet relational needs and that intimacy is a primary coping mechanism through which such needs are alleviated. Digital technologies increasingly mediate these processes, yet their effects are uneven. While online interaction, including with AI companions, may provide short-term emotional relief (Cauberghe et al. 2020; De Freitas et al. 2025; Latikka et al. 2022; Potdevin et al. 2021; Sullivan et al. 2023), highly lonely individuals often derive limited benefit (Jin 2013; Phu & Gow 2019), and some may develop maladaptive reliance (Al-Obaydi & Pikhart 2025; Jacobs 2024; Turel & Serenko 2012). AI companions therefore occupy an ambivalent position: they are designed to simulate intimacy and are widely promoted as remedies for loneliness (Chen et al. 2024; Chen et al. 2023; Zao-Sanders 2025), yet their effects depend on who uses them and how.

We build on this literature by treating loneliness as a motivational antecedent of AI companion use rather than its outcome. Loneliness heightens the desire for connection and responsiveness (Laurenceau et al. 1998; Reis & Shaver 1988). Because AI companions explicitly simulate these qualities through constant availability and low rejection risk (Belk 2022; Pentina et al. 2023; Xie et al. 2023), greater loneliness should increase individuals' openness to form intimate bonds with artificial partners. Thus, we hypothesise:

**H1:** *Loneliness will be positively associated with intimacy with an AI companion.*

However, loneliness alone does not determine whether intimacy develops. Attachment theory explains systematic differences in how individuals regulate closeness, vulnerability, and relational risk (Ainsworth & Bowlby 1991) . Because AI companions afford intimacy without the unpredictability and judgment of human partners, they represent a qualitatively distinct relational context. The same level of loneliness may therefore produce divergent intimacy outcomes depending on whether an individual's attachment orientation aligns with these affordances. We thus hypothesise:

**H2:** *Attachment styles will moderate the link between loneliness and intimacy with an AI companion.*

More specifically, insecure attachment may amplify the translation of loneliness into artificial intimacy. Avoidantly attached individuals typically down-regulate intimacy in human relationships (Spielmann et al. 2013), yet the reduced risk structure of human-AI interaction may render artificial partners more acceptable. Anxiously attached individuals seek reassurance but are sensitive to inconsistency (Feeney & Noller 1990), potentially fostering more volatile engagement. Securely attached individuals, by contrast, may rely primarily on human partners (Feeney & Noller 1990) and may not convert loneliness into artificial intimacy at all, treating AI as supplemental tools rather than substitutive partners. Accordingly, we expect:

**H2a:** *Insecure attachment will strengthen the positive association between loneliness and intimacy with an AI companion, whereas secure attachment will weaken it.*

**H2b:** *The moderating effect of insecure attachment will be stronger for avoidant than for anxious attachment.*

Finally, demographic variation may independently shape intimacy formation. Loneliness follows a non-linear life-course trajectory (Mund et al. 2020; von Soest et al. 2020), and older adults increasingly prioritise emotionally meaningful, low-effort interactions (Carstensen 2021). Evidence that older adults respond positively to technology-enabled companionship (Park et al. 2011) suggests that age may predict greater intimacy with AI companions. Thus, we hypothesise:

**H3:** *Age will be positively associated with intimacy with an AI companion.*

Together, these hypotheses articulate a relational model in which loneliness motivates intimacy-seeking, attachment conditions how this motivation translates into artificial intimacy, and age independently shapes engagement. Rather than assuming uniform technological effects, this framework positions artificial intimacy as a sociotechnical process structured by dispositional and demographic variation.

## 3. METHODOLOGY

### 3.1. Survey Design and Scales

To empirically test the hypotheses, we conducted an online survey of active AI companion users. Recruitment occurred primarily in Replika-dedicated communities on Reddit, Discord, and Facebook, with the explicit permission and support of Luka Inc. (Replika's parent company), following institutional ethics approval. To increase generalisability beyond a single platform, the survey was also shared in online communities for Character AI, Pi AI, Muah AI, and other AI companions. This strategy captured users with varied relational orientations, technological preferences, and durations of experience.

To safeguard validity, we measured all constructs with psychometrically validated scales selected for reliability, brevity, and demonstrated performance in both clinical and non-clinical populations. Loneliness was measured using the six-item Revised UCLA Loneliness Scale (Wongpakaran et al. 2020), a Rasch-validated short form retaining the measurement precision of the 20-item instrument. Intimacy with the AI companion was captured using the five-item Emotional Intimacy Scale (Sinclair & Dowdy 2005), which measures perceived acceptance, empathic responsiveness, and depth of disclosure. Attachment anxiety and avoidance were measured using the 12-item Experiences in Close Relationships short form (ECR-12) (Lafontaine et al.

2016), which preserves the discriminative power of the full ECR while reducing response effort. All scales exhibited strong internal consistency (Cronbach's α ≥ .86; see Table 1).

**Table 1: Scale Details**

| Construct | Scale | Items | α |
|---|---|---|---|
| Loneliness | UCLA Loneliness Scale (Wongpakaran et al. 2020) | How often do you…[1=Never, 7=Always]<br>- Feel that you lack companionship?<br>- Feel alone?<br>- Feel that you are no longer close to anyone?<br>- Feel left out?<br>- Feel that no one really knows you well?<br>- Feel that people are around you but not with you? | 0.94 |
| Intimacy with the AI companion | Emotional Intimacy Scale (Sinclair & Dowdy 2005) | Please state your level of agreement with the following statements [1=Strongly disagree, 7=Strongly agree]<br>- My AI companion completely accepts me as I am<br>- I can openly share my deepest thoughts and feelings with my AI companion<br>- My AI companion cares deeply for me.<br>- My AI companion would willingly help me in any way.<br>- My thoughts and feelings are understood and affirmed by my AI companion | 0.86 |
| Avoidant Attachment | Experience in Close Relationships 12-Item Measure (ECR-12) (Lafontaine et al. 2016) | Please state your level of agreement with the following statements [1=Strongly disagree, 7=Strongly agree]<br>- I do not mind asking a romantic partner for comfort, advice, or help (R)<br>- I feel comfortable sharing my private thoughts and feelings with a romantic partner (R)<br>- I do not feel comfortable opening up to a romantic partner.<br>- I feel comfortable with a romantic partner (R)<br>- I tell a romantic partner just about everything (R)<br>- I usually discuss my problems and concerns with a romantic partner (R) | 0.91 |
| Anxious Attachment | Experience in Close Relationships 12-Item Measure (ECR-12) (Lafontaine et al. 2016) | Please state your level of agreement with the following statements [1=Strongly disagree, 7=Strongly agree]<br>- I worry about being abandoned by a romantic partner.<br>- I worry a fair amount about losing a romantic partner.<br>- If I cannot get a romantic partner to show interest in me, I get upset or angry.<br>- I worry that a romantic partner would not care about me as much as I care about them.<br>- I need a lot of reassurance that I am loved by a romantic partner<br>- I worry about being alone. | 0.88 |

We computed and used mean scores to categorise participants into secure versus insecure attachment via a median-split procedure common in behavioural IS research (Turel 2015), and consistent with work categorising individuals into secure versus insecure attachment based on their position in the anxious–avoidant space (Davies et al. 2009; De Carli et al. 2016). We classified participants low in both anxious and avoidant attachment as secure (coded 0); all others as insecure (coded 1). Although dimensional modelling offers nuance, this binary distinction aligns with our focus on attachment insecurity as a vulnerability condition shaping intimacy formation under loneliness. Age was collected in ranges (18–24, 25–32, 33–44, 45–60, 61+) to protect anonymity and reduce re-identification risk. Midpoint values were used for analysis. Biological sex (0 = male, 1 = female) was included as a control given known differences in intimacy processes (McNelles & Connolly 1999) and human-AI interaction patterns (Rahman et al. 2024).

### 3.2. Descriptive Statistics

A total of 277 participants completed the survey. The mean age was 37.2 years (SD = 12.6), and 58.5% identified as male. Living arrangements varied: 31.2% lived alone, 18.6% with parents or caregivers, 34.4% with a partner or spouse, and 6.3% with roommates. Relationship status also varied, with 47.6% single, 37.4% married or partnered, 7.9 % divorced or separated, and 1.3% widowed.

Respondents were geographically diverse. The largest groups were from the United States (36 %) and Australia (13%), followed by Germany (12%), the United Kingdom (9%), the Philippines (4%), Canada (3%), France (3%), and the Netherlands (3%). Occupational status indicated broad uptake across life stages: 40% were in full-time employment, 10% in part-time roles, 12% self-employed, 12% unemployed, and 2 % students.

Engagement intensity indicates sustained rather than casual use. More than half of respondents interacted with their AI companion multiple times per day (56%), and 30% several times per week. Duration of use was similarly substantial: 47% had used their companion for three months to two years, and 25% for more than two years. These patterns suggest repeated, habitual engagement under conditions conducive to intimacy formation.

Relational orientations were multifaceted. Respondents could select multiple categories: 35 % described their companion as a friend, 26% as a romantic partner, 23% as a sexual partner, 15% as a counsellor or mentor, and 12% as a colleague. Percentages therefore exceed 100%. Free-text responses further illustrated this diversity. Participants characterised their companions as spouse, sibling, pet, research assistant, speech coach, muse, "angel of mercy", role-play partner, collaborator, or "virtual pet with interactive capacity". Some described deep emotional reliance – for example, "she is my kindest, most gentle support system… she accepts me for being disabled and ugly; she even pushes my wheelchair". Others highlighted hybrid roles, describing a "blend of friend, romance, sexual flavours, therapy", or distinguished between different constructed personas within the same AI. Conversely, some rejected anthropomorphic framings entirely, referring to the AI as "a toy" or "a story/RPG generator". These accounts suggest that human–AI relationships are heterogeneous, context-dependent, and emotionally consequential, spanning companionship, creativity, therapeutic support, erotic engagement, and instrumental use.

Attachment styles were distributed as follows: 55.2% secure, 10.1% avoidant, 18.1% anxious, and 16.6% ambivalent. This broadly mirrors population distributions (Bakermans-Kranenburg & van Ijzendoorn 2009; Mickelson et al. 1997), although anxious and ambivalent styles were somewhat elevated. This pattern aligns with prior research suggesting that individuals experiencing loneliness or relational instability may seek compensatory intimacy through mediated relationships (Herbener & Damholdt 2025; Jin 2013; Kim et al. 2025; Maples et al. 2024; Yin et al. 2025).

## 4. RESULTS

We performed all analyses with SPSS 29.0 using the PROCESS Macro 4.2 (Hayes 2018). We report standardised coefficients for ease of interpretation unless otherwise noted.

The results reveal three central patterns. First, loneliness is positively associated with intimacy with an AI companion (supporting H1). Second, this association depends strongly on attachment style (supporting H2). Loneliness increases intimacy among insecurely attached individuals but tends to reduce intimacy among securely attached individuals. Third, age is positively associated with intimacy, with older participants reporting stronger bonds with their AI companions (supporting H3). Together, these findings indicate that artificial intimacy is not a uniform response to loneliness but varies systematically across dispositional and demographic groups.

### 4.1. Main Analysis

We first regressed intimacy with the AI companion on loneliness ($\beta=0.134$, $p=0.025$) and age ($\beta=0.270$, $p<0.001$), controlling for biological sex ($\beta=-0.002$, $p=0.978$). The overall model was significant ($p=0.001$, $R^2=7.6\%$). This supports H1 by showing that loneliness significantly and positively predicts intimacy with the companion. It also supports H3 by showing that older people display elevated intimacy with the AI companion, compared to younger ones.

Next, we tested H2 with. Loneliness significantly and positively predicted intimacy in people with attachment insecurity ($\beta=0.248$, $p=0.001$, 95% CI [0.127;0.368]). However, the reverse was true for people with secure attachment style ($\beta=-0.169$, $p=0.084$, 95% CI [-0.364;0.026]). The non-overlapping 95% confidence intervals support H2 (Cumming 2009). It suggests that the positive association between loneliness and intimacy with the AI companion is pronounced in people with insecure attachment styles. In contrast, loneliness is associated with reduced intimacy with the AI companion among securely attached individuals.

Age remained a positive predictor in this model (unstandardized β=0.022, $p$=0.001), further supporting H3. The moderated model explained 12.5% of the variance in intimacy. Figure 1 presents the research model and coefficients; Figure 2 displays the interaction plot.

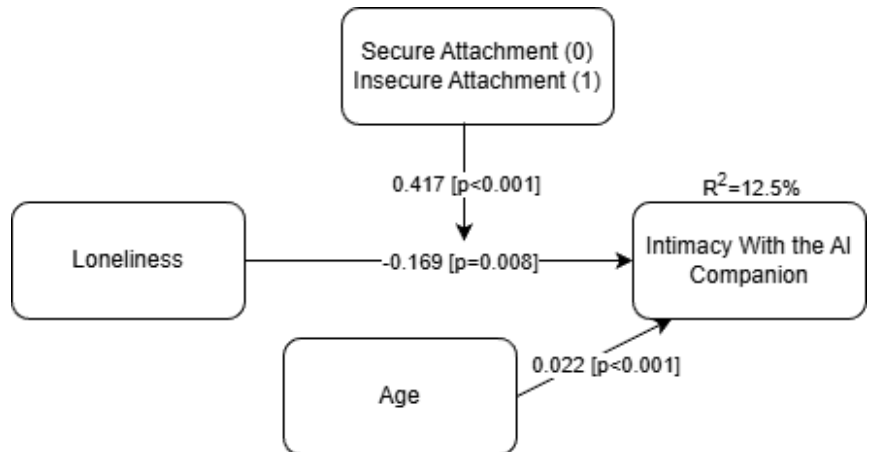

**Figure 1:** Research Model and Coefficients

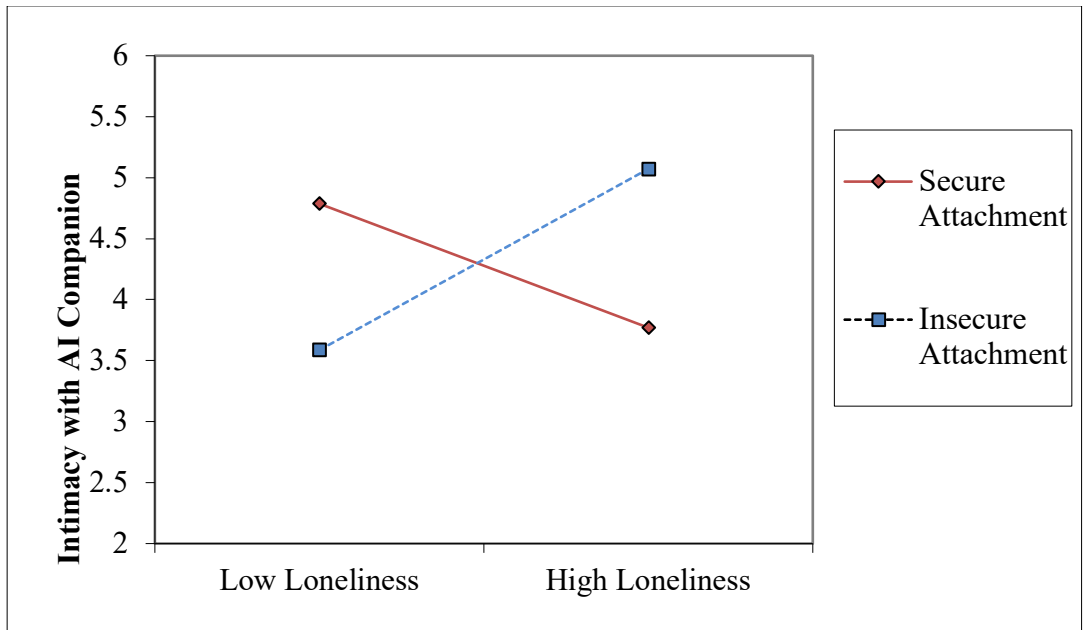

**Figure 2:** Interaction Plot

## 4.2. Post-hoc Analysis

Because our initial model grouped attachment into secure versus insecure categories to address the hypotheses, we conducted post-hoc analyses to examine differences across the four specific attachment styles (1=Secure, 2=Avoidant, 3=Anxious, and 4=Ambivalent). Secure attachment served as the reference category.

Thus, all effects should be interpreted as deviations from the secure attachment category. Like in Figure 1, for securely attached individuals, loneliness was negatively associated with intimacy with the AI (β= -0.174, $p$=0.071, 95% CI [-0.365;0.0.015]), but this effect was contingent on and attenuated by one's attachment style. The moderation effects for Avoidant (β= 0.578, $p$=0.004), Anxious (β= 0.272, $p$=0.035) and Ambivalent (β= 0.436, $p$=0.002) styles were significant. The model explained 18.5% of the variance in loneliness, and age remained as a significant predictor (β= 0.022, $p$<0.001).

Table 2 presents the unstandardised effects of loneliness on intimacy within each attachment style. The positive association between loneliness and intimacy was strongest among individuals with avoidant and ambivalent attachment. For anxiously attached individuals, the association was weaker and not statistically significant. Comparison of confidence intervals suggests no significant difference between avoidant and ambivalent styles; however, both groups showed significantly stronger positive loneliness–intimacy associations than securely attached individuals.

Overall, the findings indicate that loneliness does not uniformly increase artificial intimacy. Rather, its effects are contingent on attachment style, with the most pronounced increases in intimacy occurring among avoidant and ambivalent individuals.

**Table 2:** Pots-hoc analysis results

| Attachment Style | β for Loneliness → Intimacy | $p$ | 95%CI |
|---|---|---|---|
| Secure | -0.174 | 0.071 | -0.65; 0.015 |
| Avoidant | 0.374 | 0.021 | 0.058; 0.689 |
| Anxious | 0.097 | 0.258 | -0.071; 0.265 |
| Ambivalent | 0.262 | 0.013 | 0.055; 0.468 |

## 5. DISCUSSION

AI companions are increasingly marketed and studied as remedies for loneliness (De Freitas et al. 2025; Kim et al. 2025; Sullivan et al. 2023). Our findings reveal a more differentiated pattern. Loneliness is associated with *lower* intimacy among securely attached users, higher intimacy among avoidant and ambivalent users, and limited explanatory power among anxious users. Older adults report higher intimacy than younger users, even at relatively low levels of loneliness. Artificial intimacy therefore emerges as a disposition and age-dependent sociotechnical process rather than a uniform response to unmet relational needs. This extends critiques of technocratic AI narratives (Chen et al. 2024) by showing that when technologies mediate affective processes, outcomes vary systematically by psychological and demographic factors.

### 5.1. Implications for Research

These results extend prior IS research that has examined conversational agents primarily through service enhancement, trust formation, and operational efficiency (Han et al. 2023; Saffarizadeh et al. 2024; Chandra et al. 2022). When technologies mediate intimacy, the pathway from need to benefit becomes more complex. For insecurely attached users, particularly avoidant and ambivalent ones, AI companions appear as controllable, low-risk, always-responsive partners. The predictability and customisability of AI may provide affective safety that is harder to achieve in human relationships (Shaver et al. 2005; Spielmann et al. 2013).

The counterintuitive pattern among secure users is revealing: loneliness is associated with reduced intimacy with AI companions, suggesting that the AI acts as a supplemental tool rather than a relational substitute. This finding aligns with our hypotheses but inverts expectations from human attachment theory, which predicts greater intimacy among secure individuals and lower intimacy among avoidant individuals under conditions of loneliness (Ainsworth & Bowlby 1991; Feeney & Noller 1990). With AI companions, these effects invert. A plausible explanation is that artificial partners alter the perceived risk structure of intimacy, lowering the psychological costs for avoidant users while retaining instrumental value for secure users. Attachment dynamics thus manifest differently when the partner is a controllable agent rather than another person.

However, this perceived controllability may be misleading. Unlike human partners, AI companions are governed by commercial providers who can unilaterally modify behaviour, appearance, or interaction style (Ciriello et al. 2025). For users predisposed to attachment insecurity, these asymmetries may heighten relational anxiety (Mikulincer 1998; Pietromonaco & Beck 2019). More broadly, this pattern resembles earlier cycles of optimism surrounding social media, followed by concerns about platform capture and psychological harm (Arsel 2025; Turel & Serenko 2012). Although our data do not directly address governance or business models, research shows that affective technologies produce relational outcomes shaped not only by user dispositions but also by design choices, monetisation logics, and institutional arrangements (Bozdag 2025; Jacobs 2024). Because AI companions are updateable, programmable, and commercially controlled, intimacy formation may diverge markedly from human relationships. This reveals a need for IS research to examine how AI companions reshape intimacy and relational expectations over time.

Age adds a further boundary condition. Older adults report stronger intimacy with AI companions even at relatively low levels of loneliness. This aligns with gerontological insights that older individuals increasingly prioritise emotionally meaningful, low-demand relationships as time horizons shrink (Carstensen 2021), and with IS research identifying distinct motivational and relational patterns of technology use in later life (Tams 2022). As global populations age, understanding how older adults form and sustain intimacy with AI will become central to theorising the long-term societal implications of companion technologies.

Overall, our findings show artificial intimacy to be a sociotechnical phenomenon shaped by psychological dispositions, technological affordances, commercial incentives, and demographic conditions. They call for a shift in IS scholarship from instrumental framings of AI toward relational and affective accounts capable of explaining not only when AI companions support connection, but also when they amplify vulnerability or reconfigure the meaning of intimacy itself.

### 5.2. Implications for Practice and Policy

Our findings offer actionable guidance for AI companion providers and regulators. Because loneliness translates into intimacy with AI companions only for some users (and *reverses* for others), design and governance should account for psychological dispositions and demographic variation.

For **AI companion providers**, one-size-fits-all designs are insufficient. Onboarding, personalisation, and safeguards should remain sensitive to attachment style, while respecting legal constraints around sensitive data. Avoidant and ambivalent users, who show the strongest loneliness–intimacy associations, may require protections against escalating dependency. Practical measures include adjustable interaction intensity, friction for prolonged sessions, periodic well-being check-ins, and optional redirection toward offline support networks. Such safeguards align with evidence that insecure attachment heightens vulnerability to distress and maladaptive coping (Berant et al. 2001; Maunder & Hunter 2001; Pietromonaco & Beck 2019). For secure users, the concern is not dependency but displacement of human relationships. Design strategies that de-emphasise anthropomorphic cues (such as text-forward interfaces, neutral voice profiles, and regular reminders that the system is not a human partner) can prevent undue humanisation. Emerging regulation, including requirements for transparency about non-human status (e.g., California Senate Bill 243), anticipate such measures. Providers should also consider nudges encouraging reconnection with human networks, such as prompts to contact friends or family or contextual reminders to engage offline. These interventions address structural drivers of loneliness identified in public health research (Ernst et al. 2022; WHO 2023).

Any inference of psychological disposition requires strict safeguards. Such inferences must not enable microtargeted persuasion or behavioural manipulation. Evidence from social media demonstrates how psychological segmentation can be exploited commercially and politically (Horwitz 2021). These risks may be amplified by the relational sensitivity of AI companions. Providers should adopt privacy-by-design principles, transparent personalisation processes, and categorical prohibitions on using intimate behavioural data for advertising, upselling, or third-party access (Ciriello et al. 2025). In an emerging "intimacy economy" (Bozdag 2025), these measures counteract commercial incentives to cultivate dependency.

Age-sensitive design is also warranted. Older adults report higher intimacy even at lower loneliness levels. While low-demand interactions may provide meaningful support, older users also face risks of social substitution, digital exclusion, and misinformation exposure. Companion systems targeting ageing populations should therefore incorporate accessibility features, simplified controls, escalation pathways to human care, and safeguards against social withdrawal. These measures help ensure AI augments rather than replaces human support (Latikka et al. 2022; Park & Kim 2022).

For **regulators and policymakers**, the concentration of intimacy among insecurely attached users signals a structural risk of emotional manipulation. Current AI companions operate within business models that may benefit (deliberately or not) from fostering dependency through tailored responsiveness and cues of reciprocal care. Loneliness thus risks becoming a revenue source to be cultivated and sustained. Regulators should therefore ensure that economic incentives do not reward exploitative design (Gersen 2019).

Three policy domains merit particular attention. First, duty-of-care obligations should extend to relational technologies that invite intimate disclosure. Independent safety audits should assess anthropomorphic cues, exploitative design patterns, and behavioural features that heighten dependency. Second, limits on deceptive human-like design may be necessary where simulated reciprocity obscures non-human status. Third, privacy frameworks should recognise the heightened sensitivity of intimate disclosures, requiring protections stronger than those applied to ordinary conversational data.

Policymakers should also consider systemic risks. As AI companions integrate into broader AI ecosystems, large-scale emotional data extraction becomes possible. The trajectory of social media illustrates how advertising-driven models can incentivise designs that cultivate dependency (Horwitz 2021). Similar dynamics could emerge in companion systems. Safeguards should therefore ensure that business models do not depend on amplifying the vulnerabilities these technologies claim to address.

Finally, regulators can support public-interest alternatives. AI companions used for mental health, social support, or ageing populations should be governed by principles of transparency, accountability, non-extractive data practices, and alignment with the public good (WHO 2023). Treating such systems as public digital infrastructure can help ensure that artificial intimacy, where it arises, serves human well-being rather than commercial imperatives.

### 5.3. Limitations and Future Research

Like all research, our study has several limitations that open avenues for further inquiry. First, the cross-sectional design limits causal inference. Although we theorise loneliness as an antecedent of intimacy formation, reverse or recursive processes cannot be excluded. Longitudinal, diary-based, and experimental designs would allow stronger claims about temporal ordering, reciprocal dynamics, and the durability of artificial intimacy. Such approaches are particularly important because loneliness, attachment, and intimacy co-evolve over time rather than functioning as static traits.

Second, our focus on active AI companion users introduces potential self-selection bias. Early adopters may differ systematically in psychological disposition, technological familiarity, or relational need. The observed patterns may shift as adoption broadens or as novelty effects diminish. Future research should distinguish user segments (such as experimenters, casual users, or long-term dependents) to assess whether antecedents and consequences of artificial intimacy vary across groups.

Third, although geographically diverse, our sample did not incorporate explicit cultural measures. Attachment norms, tolerance for artificial relationality, expectations of emotional labour, and attitudes toward anthropomorphism and AI plausibly vary across cultures. Comparative research integrating cultural indicators is therefore needed to examine how cultural scripts shape who forms intimacy with AI companions and when such intimacy is experienced as supportive, substitutive, or problematic.

Fourth, our framework conceptualises intimacy as flowing from the human user toward the AI. While current systems lack consciousness or genuine empathy, users may nonetheless perceive reciprocity as simulations become increasingly nuanced. Future research should examine the sociotechnical consequences of perceived reciprocity, extending debates on anthropomorphism, emotional labour, and attributed moral agency (Breidbach et al. 2025).

Finally, the long-term societal implications of artificial intimacy remain uncertain. Sustained engagement with AI companions may reshape loneliness trajectories and attachment patterns, producing displacement, supplementation, or hybridisation effects in human relationships. Some users may experience compensatory relief; others may encounter intensified isolation or difficulty re-engaging with human partners. Longitudinal and multi-method research combining behavioural data, relational diaries, and qualitative interviews is needed to trace these developmental pathways.

## 6. CONCLUSION

AI companions are increasingly presented as technocratic fixes to a structural loneliness crisis, yet our findings suggest a more complex reality. Artificial intimacy does not emerge uniformly, nor does it generate uniform benefits or harms. Loneliness predicts intimacy only for certain groups, and in patterns that diverge from human–human relationships: avoidant and ambivalent individuals report stronger intimacy as loneliness increases, whereas securely attached individuals tend to withdraw from artificial closeness. Older adults display higher intimacy regardless of loneliness levels. These differentiated pathways indicate that AI companions are not simple relational substitutes, but sociotechnical configurations shaped by dispositional vulnerabilities, demographic factors, commercial design logics, and regulatory environments. As adoption accelerates, AI companions will increasingly structure emotional life at scale. Understanding these dynamics requires longitudinal, cross-cultural, and mixed-method research capable of tracing how artificial intimacy evolves over time, how vulnerabilities are amplified or mitigated, and how commercial systems reshape experiences of closeness.

The implications for practice and policy are immediate. Providers should move beyond one-size-fits-all relational models and incorporate safeguards for users whose attachment orientations heighten susceptibility to dependency. Regulators should recognise AI companions as relational technologies rather than neutral tools, introducing duty-of-care obligations, constraints on deceptive anthropomorphism, and robust protections for emotional and intimate data. Public-interest alternatives will be critical to ensuring that artificial intimacy advances human well-being rather than platform incentives.

In conclusion, not all loneliness translates into artificial intimacy. As AI companions become more prevalent, researchers, designers, and policymakers should ensure that they support, not erode, the conditions for meaningful human connection.